# RADIOACTIVE ISOTOPES REVEAL A NON SLUGGISH KINETICS OF GRAIN BOUNDARY DIFFUSION IN HIGH ENTROPY ALLOYS

M. Vaidya[a,b], K.G. Pradeep[c], B.S. Murty[b], G. Wilde[a,d], S.V. Divinski[a,e]

[a]*Institute of Materials Physics, University of Münster, Germany*
[b]*Department of Metallurgical & Materials Engineering, Indian Institute of Technology Madras, Chennai - India*
[c]*Materials Chemistry, RWTH Aachen University, Kopernikusstr.10, 52074, Aachen, Germany*
[d]*Nanjing University of Science & Technology, Herbert Gleiter Institute of Nanoscience, Jiangsu, P. R. China*
[e]*Samara National Research University, Moskovskoye Shosse 34, Samara, 443086, Russia*

*High entropy alloys (HEAs) have emerged as a new class of multicomponent materials, which have potential for high temperature applications. Phase stability and creep deformation, two key selection criteria for high temperature materials, are predominantly influenced by the diffusion of constituent elements along the grain boundaries (GBs). For the first time, GB diffusion of Ni in chemically homogeneous CoCrFeNi and CoCrFeMnNi HEAs is measured by radiotracer analysis using the $^{63}$Ni isotope. Atom probe tomography confirmed the absence of elemental segregation at GBs that allowed reliable estimation of the GB width to be about 0.5 nm. Our GB diffusion measurements prove that a mere increase in number of constituent elements does not lower the diffusion rates in HEAs, but the nature of added constituents plays a more decisive role. The GB energies in both HEAs are estimated at about 0.8–0.9 J/m$^2$, they are found to increase significantly with temperature and the effect is more pronounced for the CoCrFeMnNi alloy.*

In the 21$^{st}$ century, the insistent campaign for enhancing energy efficiency in power generation and aerospace technology renders the development of high performance materials the foremost goal of materials scientists. For instance, the jet engine operating temperature has direct impact on its service life-time, efficiency and carbon emission. Materials performance at high temperatures is primarily limited by the onset of plastic deformation process called creep [1]. One of the critical factors that influence the creep deformation in materials is the presence of two-dimensional defects known as grain boundaries. A grain boundary (GB) is an interface present between two grains developed during crystallization, growth or during processing of materials by thermo-mechanical treatments. While crystalline grains have regular periodic arrangements of atoms, grain boundaries reveal specific structures which are generally characterized by an increased free volume [2] and hence can provide easier pathways for atomistic diffusion and sliding during creep deformation. In addition, grain boundaries being high energy regions reduce the phase stability by acting as heterogeneous nucleation sites for precipitation of new secondary phases. Thermo-mechanical treatments can be utilized to tailor the grain boundary characteristics and hence the microstructure of materials via recrystallization, grain growth, segregation, which influence the mechanical properties. Similarly, processes such as sintering and stress corrosion cracking are also influenced by the existence of grain boundaries [2]. Hence, defects present in metallic materials both by varied types and fractions determine the resulting mechanical properties. A vital property, involving a specific defect type namely the grain boundary, is the atomic transport or diffusion along the boundary. It is the rate-limiting step in several key deformation processes and phase transformations. The most direct and reliable experimental technique to determine the GB diffusion is the radiotracer diffusion method [3] that allows measuring self-diffusion. Use of a radioactive isotope (in ppm concentrations) of an alloy constituent ensures that the measurements are performed in a chemically identical environment without additional chemical driving forces for atomic mobility. Thus, an accurate quantification of GB energy is feasible and is fundamental for the understanding and hence development of new structural materials targeted for high temperature applications. This correlation has been schematically represented in Fig. 1.

A new class of multi-component alloys termed high entropy alloys (HEAs) have emerged over the last decade which promises to present attractive mechanical properties including uncompromised strength-ductility combinations [4]. These alloys feature the unique characteristics of possessing multi-component constituent elements in equimolar or near equimolar proportions. Accordingly, the maximized configurational entropy state is believed to provide enhanced phase stability and the presence of a multi-element matrix presumably decelerates the diffusion rates [5]. HEAs, by virtue of the above mentioned characteristics, have shown some fascinating properties like high damage tolerance [6], enhanced fracture toughness [7], high strength combined with ductility [4], superior oxidation [8] and corrosion resistance [9]. Consequently, HEAs are being constantly explored as high temperature materials.

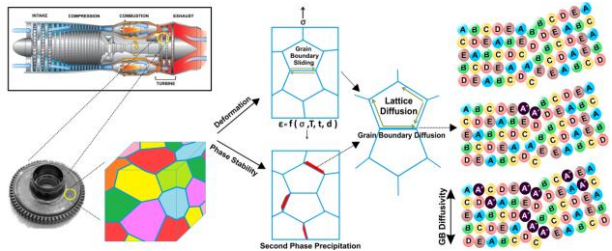

**Fig. 1:** Importance of GB diffusion in potential applications. Grain boundaries are pivotal in determining creep deformation and phase stability, which directly impacts the service life of the component material.

In this study, we report for the first time the GB diffusivities determined by employing the radiotracer method (using the $^{63}$Ni isotope) in two single phase FCC structured HEAs, namely CoCrFeNi and CoCrFeMnNi. Chemically homogenous alloys, with proven phase stability [10], have been prepared to ensure zero influence of secondary phases or segregated impurities. An extended temperature interval (673 – 1173 K) is selected to warrant the consistency of the obtained values and also to determine the GB width. It may be emphasized that, until recently the diffusion studies in HEAs have been scarce and limited to the estimation of lattice (or bulk) inter-diffusion coefficients using the diffusion couple method [11–14]. It is important to note that it is indeed difficult to differentiate between the dominance of lattice or grain boundary diffusion contributions in an interdiffusion approach. Our previous work utilized the radiotracer technique to overcome this fundamental difficulty and to provide true bulk diffusion parameters in CoCrFeNi and CoCrFeMnNi HEAs [10].

Thus, GB diffusion data for HEAs are completely missing so far. The GB diffusion coefficients reported in the present study are therefore essential to guide the selection and design of HEAs as high performance materials.

**Methods**

The CoCrFeNi and CoCrFeMnNi alloys studied in this work were processed by arc melting using solid pieces (99.99 wt.% purity) of Co, Cr, Fe, Mn and Ni in equiatomic proportions. The melting chamber was first evacuated to $10^{-5}$ mbar pressure and then purged with highly purified Ar to remove any excess oxygen. Chemical homogeneity was ensured by re-melting each alloy 4-5 times. The melted alloy buttons were further homogenized at 1473 K for 50 h. The structure of both alloys was confirmed by X-ray diffraction (XRD) using PANalytical X'Pert PRO diffractometer using Cu $K_\alpha$ radiation. The microstructure of CoCrFeNi and CoCrFeMnNi alloys was investigated with scanning electron microscope (SEM, FEI Inspect F) equipped with electron back-scattered diffraction (EBSD) and energy dispersive spectroscopy (EDS).

The local chemical homogeneity of the alloys was further characterized using local electrode atom probe tomography (LEAP 4000X HR) provided by CAMECA instruments. Site specific tips containing GBs for atom probe tomography (APT) were prepared using a FEI Helios Nanolab 660 dual beam focussed ion beam following the procedures described in [15,16]. APT measurements were performed with the tips maintained at 60 K applying laser pulses at 250 kHz frequency and 20 pJ laser energy. Data reconstruction and analysis was performed with the IVAS 3.6.10 software.

Cylindrical samples of 8 mm in diameter and 1 mm in thickness were cut and mechanically polished. A highly diluted acidic solution of $^{63}$Ni isotope (68 keV β-decays, half-life of 100 years) was deposited on the sample surface and allowed to dry. The samples were sealed in quartz tubes under purified (5N) Ar atmosphere and subjected to diffusion annealing treatments at 673 K, 723 K, 873 K, 1073 K and 1173 K for different times. A Ni/NiCr thermocouple (type K) was used to measure the sample temperature, which was controlled within ±1 K. The subsequent steps involved reduction of the sample diameter by 1 mm to eliminate artifacts caused by surface and lateral diffusion.

A high-precision grinding machine was used to determine the penetration profiles by a serial sectioning technique. Since the diffusion depths at low temperatures are very small, the thickness of ground sections was kept as small as possible, about 0.3 μm, allowing nevertheless a reliable determination of the concentration profiles. The intensity of β-decays was determined by a Liquid Scintillation Counter TRI CARB 2910TR and the counting times were suitably chosen to keep the statistical uncertainty of the measurements below 2%.

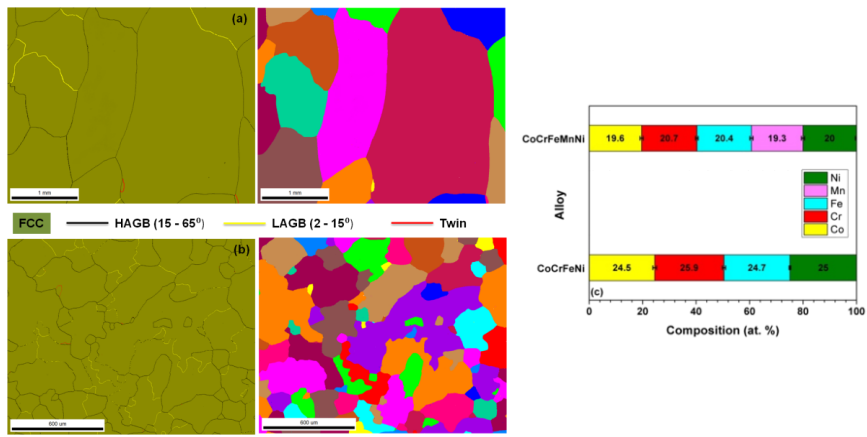

**Fig. 2:** EBSD phase maps and unique color grain maps for a) CoCrFeNi and b) CoCrFeMnNi. Single phase FCC is indexed and different types of GBs are color coded. HAGBs are clearly predominant. c) An equiatomic composition is obtained in both alloys.

## Results and Discussion

*Microstructure analysis and chemical homogeneity at GBs*

A thorough knowledge of the alloy structure (nature of phases, microstructure and potential element segregation) is essential to assess the diffusion along GBs. Both CoCrFeNi and CoCrFeMnNi alloys possess single phase FCC structures. Figures 2a and 2b show EBSD phase and grain maps for the CoCrFeNi and CoCrFeMnNi alloys, respectively. EBSD indexing showed simple FCC phase for both the alloys as indicated by green color in the phase maps. High-angle grain boundaries (HAGBs) are highlighted in black color, low-angle grain boundaries (LAGBs) are marked by yellow line segments and twins are indicated in red. Clearly, the majority of grain boundaries are HAGBs. Although CoCrFeMnNi shows a increased fraction (~8% more) of LAGBs, it is not expected to affect the present GB diffusion measurements, since the LAGBs reveal typically significantly lower diffusion rates compared to those for the HAGBs [2], although a corresponding contribution can be observed in dedicated experiments [17]. The unique color grain maps demonstrate that grains are equiaxed, with average grain sizes in excess of 250 μm, typical for materials in commercial applications. The estimated bulk composition of the alloys is presented in Fig. 2c which indicates the presence of equiatomic concentrations of the constituent elements.

For reliable interpretation of the GB diffusion measurements it is also imperative to characterize the local composition near the grain boundaries with respect to potential GB segregation. The CoCrFeNi and CoCrFeMnNi HEAs were studied using APT with a focus on GB segregation, if any, and the typical results are shown in Fig. 3. In the case of CoCrFeNi, the investigated volume includes a single random HAGB (Fig. 3a). In addition to the main alloying elements, trace impurity carbon atoms (0.0087 at.%) were detected and their distribution (insert) does not indicate any GB segregation. The binomial frequency distribution analysis in Fig. 3a, confirm that all elements are uniformly distributed and the composition is equiatomic even at the particular GB.

In the case of CoCrFeMnNi, a triple junction with the corresponding joining GBs was selected for analysis. A homogenous distribution of the constituent elements plus carbon (shown only for 0.0173 at.% carbon atoms in the insert) across the triple junction and a nearly equiatomic composition both within the grains and at the triple junction as well as at the grain boundaries is substantiated in Fig. 3b for the quinary alloy. The homogeneity of alloys is further confirmed by the three dimensional elemental maps of each of the constituent elements in CoCrFeNi (Fig. 4a) and CoCrFeMnNi (Fig. 4b).

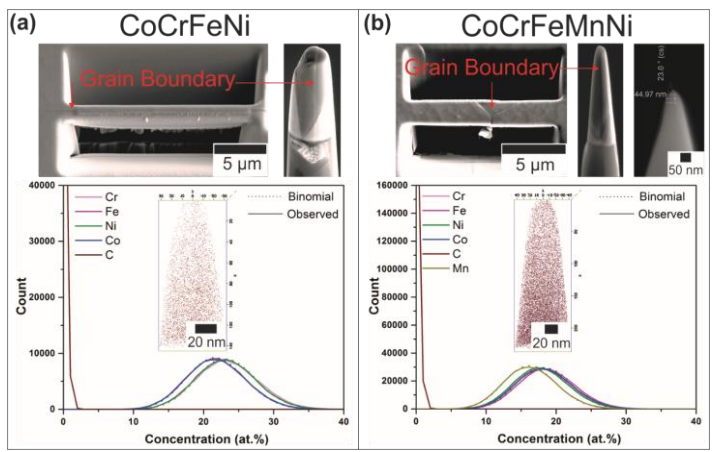

**Fig. 3:** APT results for (a) CoCrFeNi and (b) CoCrFeMnNi alloys. The site-specific grain boundary samples for each alloy were prepared using FIB. The frequency distributions with 100 ions per bin for all main elements and C are plotted. The inserts show distributions of carbon atoms within the tips.

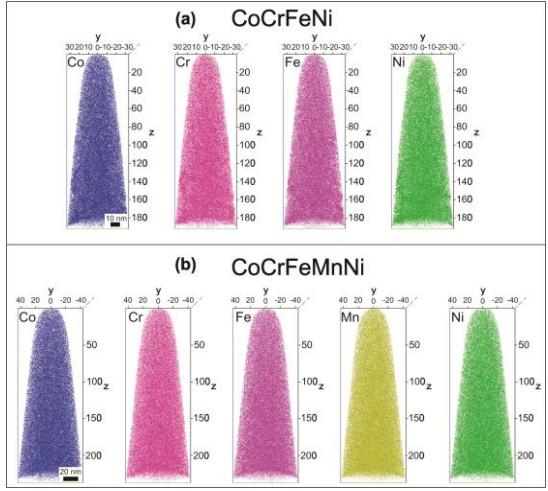

**Fig. 4:** Elemental maps of constituents in a) CoCrFeNi and b) CoCrFeMnNi alloys obtained using atom probe tomography. A homogeneous distribution of all the elements is further substantiated.

The stability of the FCC solid solution during long term annealing at higher temperatures (1073 K, 31 days) was established for both CoCrFeNi and CoCrFeMnNi in our previous study [10]. Differential thermal analysis (DTA) was employed in the earlier study [10] to determine the melting point ($T_m$, taken as melting peak temperature during DTA) of both the alloys. The measurements confirmed the stability of the single phase structure up to the melting point, too. This is consistent with the literature studies [18,19], in which it was, e.g., shown that the CoCrFeMnNi alloy reveals a single phase FCC structure and it is retained even after prolonged heat treatments at high temperatures (1173 K, 500 days [19]). A phase decomposition in these alloys has been reported [18,20] to occur when either the microstructure is fine grained and/or extremely long annealing times at moderate temperatures are used, neither of which is the case in the present study.

Thus, the CoCrFeNi and CoCrFeMnNi alloys selected for GB diffusion measurements possess stable single phase FCC structures, homogenous distributions of elements, large fractions of HAGBs and no segregation at the GBs. Hence, Ni diffusion along general HAGBs can reliably be measured in the present study.

*Penetration profiles of Ni GB diffusion in HEAs*

The contribution of GBs to atomic transport in materials varies according to the temperature, time of exposure and grain size. Harrison [21] classified the kinetic regimes of grain boundary diffusion as A-, B- or C- type. The A-type regime in a polycrystalline material holds at high temperatures and long annealing times, when the parameter $\Lambda$ ($\Lambda = d/\sqrt{D_v t}$) is smaller than 0.3 [22]. Here $D_v$ is the bulk diffusion coefficient (determined for the HEAs in question in our previous study [10]), $d$ is the grain size and $t$ is the annealing time. In the present measurements, the bulk diffusion length, $\sqrt{D_v t}$, is significantly smaller than the grain size (see Table 1) and the relation $\Lambda > 3$ holds. For example, the maximum value of $\sqrt{D_v t}$ is about 6.4 µm (for CoCrFeMnNi at 1173 K), which is less than 1/20th of the grain size ($d > 250$ µm). Thus, the present measurements definitely belong to the B- or C-type kinetic regimes depending on temperature and diffusion time. The concentration profiles of Ni GB diffusion in CoCrFeMnNi presented in Fig. 5a and 5b exemplify the diffusion contributions observed in the C- and B-type kinetic regimes, respectively, and similar profiles are observed at all temperatures for both quaternary and quinary alloys. In order to demarcate the B- and C-type kinetic regimes, the value of the Le Claire parameter $\alpha$,

$$\alpha = \frac{s\delta}{2\sqrt{D_v t}} \quad (1)$$

is decisive [23]. Here, $s$ is the segregation factor and $\delta$ is the diffusional GB width (measured to be equal to about 0.5 nm in FCC metals [3,24]). Considering the absence of elemental segregation at the GBs, as it is confirmed by the present APT study, $s$ can safely be assumed to be equal to 1 for Ni diffusion in both, CoCrFeNi and CoCrFeMnNi alloys.

*The C-type regime of GB diffusion*

The C-type kinetic regime conditions hold when $\alpha > 1$ [22]. In this regime, which corresponds typically to lower temperatures and shorter annealing times, the tracer diffusion into the bulk is negligible and atomic transport occurs primarily along the grain boundaries. In the present case, based on the value of the parameter α (see Table 1), the C-type measurements are carried out at 673 K for the quaternary alloy and below 723 K for the quinary alloy. The Gaussian solution can be applied and the logarithm of the tracer concentration, $\bar{c}$, is proportional to the penetration depth squared, $x^2$. Thus the grain boundary diffusion coefficient, $D_{gb}$, can be determined as

$$D_{gb} = \frac{1}{4t}\left(-\frac{\partial \ln \bar{c}}{\partial x^2}\right)^{-1} \quad (2)$$

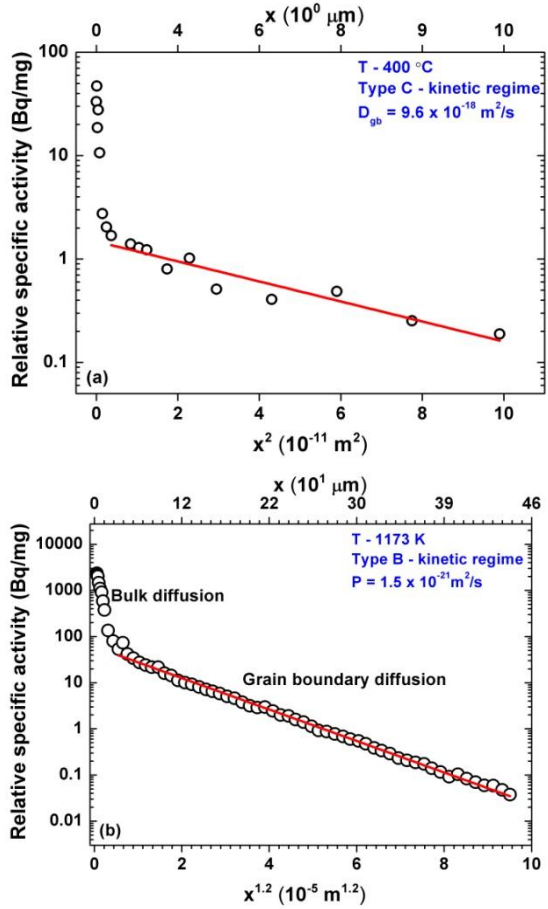

**Fig 5:** Penetration profiles of Ni GB diffusion in CoCrFeMnNi measured at (a) 673 K in the C kinetic regime and (b) 1173 K in the B kinetic regime. The derived values of $D_{gb}$ and P from the respective profiles are indicated.

As an example, Figure 5a shows a penetration profile of Ni diffusion measured in the CoCrFeMnNi alloy in the C-type kinetic regime at 673 K. After an initial drop of the tracer activity by more than two orders of magnitude (related to grinding-in effects), a prolonged tail corresponding to GB diffusion is measured and it is almost linear with respect to $x^2$. The penetration depth of bulk diffusion at 673 K is a fraction of a nanometer (see Table 1) and the near-surface data points of the corresponding penetration profiles represent a superposition of the grinding-in effects and a potential diffusion contribution of LAGBs or dislocations [25]. These points will not be analyzed further in the present paper, since a detailed investigation is required. However, the linear branches of the penetration profiles allow a reliable determination of the GB diffusion coefficients, $D_{gb}$ (see Table 1).

*The B-type regime of GB diffusion*

At higher temperatures, the contribution of bulk diffusion is important and the B-type kinetic regime holds if α < 0.1. In this case, a GB diffusion-related branch of the profile is characterized by a linear dependence of the logarithm of the tracer concentration vs. $x^{6/5}$ according to Le-Claire's analysis [23] of Suzuoka's exact solution [26]. The corresponding slope yields the value of the GB triple product $P$, $P = s·\delta·D_{gb}$,

$$P = 1.308\sqrt{\frac{D_v}{t}}\left(-\frac{\partial \ln \bar{c}}{\partial x^{6/5}}\right)^{-5/3} \quad (3)$$

As an example, the concentration profile for Ni GB diffusion in CoCrFeMnNi measured at 1173 K is shown in Fig. 5b. Both bulk diffusion and GB diffusion branches are clearly seen and the GB diffusion conditions correspond to the B-type regime. The relevant experimental parameters and the determined triple product values are listed in Table 1.

Table 1 shows the values of the GB diffusion parameters α, β, and the bulk diffusion length, $\sqrt{D_v t}$. Clearly, the measurements at temperatures above 723 K fall into the B-type kinetic regime for both alloys and the experiments at $T < 723$ K were performed in the C-type kinetic regime. However, a single measurement at 723 K in the CoCrFeNi alloy corresponds to α = 0.9 which belongs to the transition BC kinetic regime. The analysis is outlined in next Section.

The GB tail is more extended at higher temperature, i.e.1173 K, than at 673 K in accordance with the enhanced diffusion rates at elevated temperatures. The diffusion profile presented in Fig. 5b is typical for the B-type kinetic regime [22], showing an increased lattice diffusion contribution. A linear combination of two appropriate exponential functions, $\exp(-q_1 x^2)$ and $\exp(-q_2 x^{1.2})$ (with $q_1$ and $q_2$ being fitting parameters), has been used to fit these penetration profiles and to determine the volume diffusion coefficients and the triple product values, $P$, respectively (Note that the Ni volume diffusion coefficients were reported previously [10]). In a few cases, the bulk diffusion branch was found to be better approximated by an error function solution [10], however this fact does not affect the derived values of GB diffusion parameters.

**Table 1:** Parameters of GB diffusion measurements. The uncertainties of the determined values of $P$ and $D_{gb}$ do not exceed 20%.

| CoCrFeNi | | | | | | | |
|---|---|---|---|---|---|---|---|
| T (K) | t ($10^5$ s) | $\sqrt{D_v t}$ (nm) | $D_{gb}$ (m$^2$s$^{-1}$) | P (m$^3$s$^{-1}$) | α | β | Kinetic regime |
| 673 | 5.78 | 0.1 | $9.6 \times 10^{-18}$ | ---- | 2.4 | ---- | C |
| 723 | 1.76 | 0.3 | $6.35 \times 10^{-17a}$ | $4.5 \times 10^{-26a}$ | 0.9 | $1.6 \times 10^8$ | Transition BC |
| 873 | 6.05 | 21 | ---- | $2.6 \times 10^{-24}$ | $1.2 \times 10^{-2}$ | $8.2 \times 10^4$ | B |
| 1073 | 11.8 | 808 | ---- | $1.5 \times 10^{-22}$ | $2.1 \times 10^{-4}$ | $1.3 \times 10^3$ | B |
| 1173 | 4.03 | 1617 | ---- | $1.5 \times 10^{-21}$ | $1.5 \times 10^{-4}$ | $7.7 \times 10^1$ | B |
| **CoCrFeMnNi** | | | | | | | |
| 673 | 5.78 | 0.03 | $5.7 \times 10^{-18}$ | ---- | 8.2 | $4.0 \times 10^9$ | C |
| 723 | 1.76 | 0.11 | $1.3 \times 10^{-17}$ | ---- | 2.3 | $2.4 \times 10^8$ | C |
| 873 | 6.05 | 15 | ---- | $4.0 \times 10^{-24}$ | $1.6 \times 10^{-2}$ | $1.8 \times 10^5$ | B |
| 1073 | 29.3 | 1698 | ---- | $3.6 \times 10^{-21}$ | $1.5 \times 10^{-4}$ | $1.1 \times 10^3$ | B |
| 1173 | 22.5 | 6364 | ---- | $2.6 \times 10^{-20}$ | $3.9 \times 10^{-5}$ | $1.1 \times 10^2$ | B |

[a] These are uncorrected values. For the corrected ones, see the text.

**Table 2:** Arrhenius parameters for Ni GB diffusion in HEAs and other FCC matrices. $Q$ is the activation energy and $P_0$ the pre-exponential factor for GB diffusion.

| Matrix | $T_m$ (K) | $Q_{gb}$ (kJ/mol) | $P_0$ ($10^{-12}$ m$^3$s$^{-1}$) | $D_{0gb}$ ($10^{-2}$ m$^2$s$^{-1}$) $= P_0/\delta$ | $Q_{gb}^* = Q_{gb}/RT_m$ | $Q_v$ (kJ/mol) | $Q_v^* = Q_v/RT_m$ | Ref.[a] |
|---|---|---|---|---|---|---|---|---|
| Ni | 1728 | 128 | 0.0088 | 0.00176 | 8.9 | 278 | 19.3 | [27][33] |
| Fe-40Ni | 1723 | 157 | 0.113 | 0.026 | 11.0 | 303 | 21.2 | [32][34] |
| CoCrFeNi | 1717 | 158±7 | $0.011^{+0.019}_{-0.007}$ | $0.0022^{+0.005}_{-0.002}$ | 11.3 | 258 | 18.1 | Present work, [10] |
| CoCrFeMnNi | 1607 | 221±14 | $142^{+915}_{-123}$ | $28^{+190}_{-25}$ | 16.5 | 304 | 22.7 | Present work, [10] |

[a] The order of references indicate the source for $Q_{gb}$ and $Q_v$ values respectively.

### The transition BC kinetic regime

Szabo et al. [45] have shown that if a penetration profile measured in the transition BC kinetic regime is nevertheless analyzed according to the B- or C-type of the kinetic regimes, the determined values of $P$ or $D_{gb}$, respectively, will be underestimated. The authors suggested a simple procedure to correct the experimental values and we will use it here as described in Ref. [22]:

Taking into account that the value of α is almost unity in this experiment at 723 K, the following steps are taken:

1. The concentration profile is re-plotted against the reduced penetration depth $\alpha w^{4/5}$, where $w = \dfrac{x}{\sqrt{\delta D_{gb}}}\left(\dfrac{4D_v}{t}\right)^{1/4}$. The evaluated value of $D_{gb}$ is used here and $y$ is the penetration depth.

2. Then the average value of this parameter $\alpha w^{4/5}$ is determined and the correction factor, $D_{gb}^{exp}/D_{gb}^{theor}$, could be found from the graphical dependence shown in Ref. [22]. Here $D_{gb}^{theor}$ is the true value of the GB diffusion coefficient.

Having applied this procedure, the correction factor is found to be 0.45 and the "true" value of the Ni GB diffusion coefficient at 723 K is then $1.22 \times 10^{-16}$ m$^2$/s.

### Temperature dependence of Ni GB diffusion in HEAs

Figure 6a shows the Arrhenius plots for the determined GB diffusion parameters $P$ and $D_{gb}$ for both alloys, with the latter being multiplied by the GB width $\delta$ taken as 0.5 nm. As a result, almost linear temperature dependencies are observed. This fact allows concluding that the product $s \cdot \delta = P/D_{gb} \approx 0.5$ nm in both HEAs. This conclusion has two consequences:

- The segregation factor of Ni at grain boundaries in both HEAs is about unity, which completely agrees with the direct APT observations, Figs. 3 and 4;
- The diffusional GB width $\delta$ is about 0.5 nm in the high-entropy alloys.

The latter estimate corresponds well to the previous results on the GB width in FCC metals and alloys, including Ni [24,27], and Fe-Ni alloys [28], which are relevant for the present study. Further, it also indicates that "severe lattice distortion" (a "core effect" in HEAs), which has recently been shown to be small in bulk CoCrFeMnNi [29,30], is small at GBs too.

Figure 6a substantiates again the absence of a hypothetical general diffusion retardation in HEAs, in this particular case with respect to GB diffusion transport and similar to the case of bulk diffusion considered in Ref. [10]. An increase of the number of alloying components in HEAs does not automatically result in sluggish diffusion as it was speculated in early investigations [11]. At higher temperatures, about 800 K, the Ni GB diffusion rate in a quinary HEA is even higher than that in the quaternary alloy, whereas the relationship is reversed at lower temperatures.

In the present work, Ni GB diffusion profiles were measured in an extended temperature interval, including relatively high temperatures, at which a strong contribution of short-circuit diffusion in relatively coarse polycrystal may be somewhat unexpected. Indeed, after Fisher [31], the GB diffusion contribution to the total layer tracer concentration, $\bar{c}$, which is determined in a sectioning experiment, is (see also Ref. [22]),

$$c_0 \frac{\delta}{d}\left[1 + \frac{2}{\sqrt{\pi}} \frac{2\sqrt{D_v t}}{\delta}\right] \exp\left(-\frac{x}{L_{gb}}\right) \qquad (4)$$

where $c_0$ is the surface tracer concentration and $L_{gb}$ is the GB diffusion length,

$$L_{gb} = \left(\frac{\pi t}{4 D_v}\right)^{1/4} \sqrt{\delta D_{gb}} \qquad (5)$$

From Eq. (4) one immediately recognizes that the interface contribution to the layer concentration $\bar{c}$ scales as $\frac{\delta}{d}$ in the C-type regime and as $\frac{\sqrt{D_v t}}{d}$ in the B-type regime. For a comparison, in Fig. 7 the penetration profiles measured for Ni diffusion in CoCrFeMnNi at 1173 K (the present work) and for Ni diffusion in 99.99wt.% pure Ni at 1100 K [24] are compared. The Fisher coordinates are used and the measured tracer activity is normalized by the bulk diffusion length, $\sqrt{D_v t}$, whereas the penetration depth $x$ is divided by the GB diffusion length $L_{gb}$, see Eq. (5). Since we are applying similar amounts of the tracer material in our GB diffusion experiments, about 10 kBq, the penetration profiles can directly be compared. Similar contributions of short-circuit diffusion in HEAs with respect to that in pure Ni are seen in view of similarly large grain sizes (>250μm). We conclude that the grain boundaries in the CoCrFeNi and CoCrFeMnNi HEAs are similar effective short-circuits for Ni atoms as those in pure Ni despite the GB diffusion rates in the HEA are enhanced at high temperatures and retarded at lower temperatures below 1000 K as compared to pure Ni.

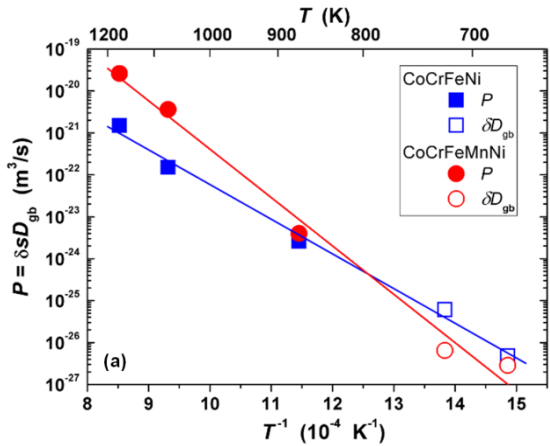 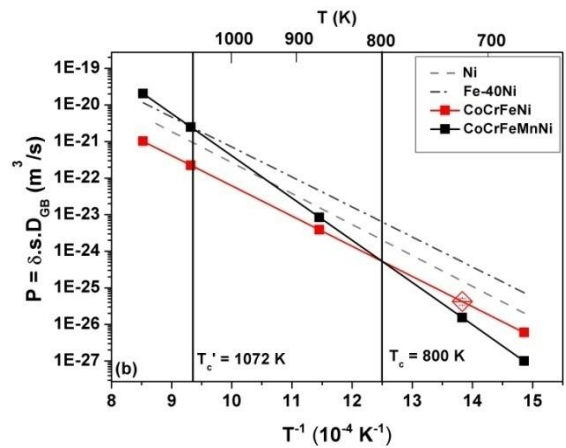

**Fig. 6:** (a) Arrhenius plots of Ni GB diffusion in CoCrFeNi (circles) and CoCrFeMnNi (squares) measured in B- (the $P$ values, solid symbols) and C-type (the $\delta \cdot D_{gb}$ values, open symbols) kinetic regimes. The values of the Ni GB diffusion coefficients are multiplied by the GB width $\delta$ taken as 0.5 nm. (b) Comparison of GB diffusivities (triple products) of Ni in CoCrFeNi and CoCrFeMnNi alloys with pure Ni [27] and Fe-Ni alloy [32] against the inverse of the absolute temperature.

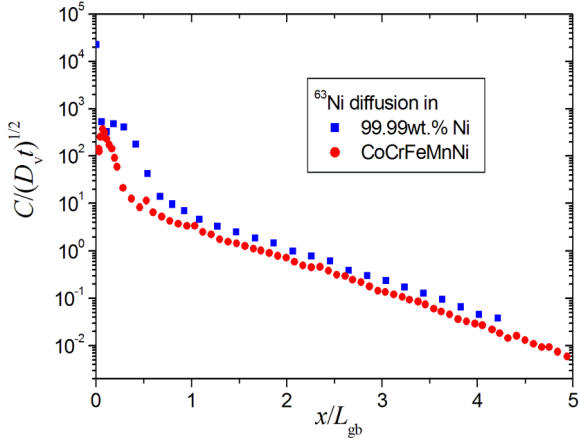

**Fig. 7:** Penetration profiles measured for Ni GB diffusion in polycrystalline CoCrFeMnNi at 1173 K (circles, present work) and in 99.99wt.% pure Ni at 1100 K [24] (squares) represented in the normalized coordinates $C/\sqrt{D_v t}$ vs. $x/L_{gb}$ referring to the Fisher solution, see text.

In Fig. 6b the measured triple products of GB diffusion in the HEAs under consideration are compared with the literature values for Ni GB self-diffusion in pure Ni [27] and in a binary Fe-Ni alloy [32]. The variation of Ni GB diffusivity with respect to the inverse of the respective homologous temperatures ($T/T_m$) also follows a similar trend (Fig. 8). Ni GB diffusion in CoCrFeNi is slower just by an order of magnitude when compared to pure Ni and the binary alloy, while CoCrFeMnNi showed decelerated diffusivities only above a cross-over temperature (discussed later).

The GB diffusion parameters ($Q_{gb}$, $P_0$, $Q_{gb}^* = Q_{gb}/RT_m$) along with the bulk diffusion parameters ($Q_v$ and $Q_v^* = Q_v/RT_m$) of Ni in all the four FCC matrices (pure Ni [33], Fe-Ni [34], CoCrFeNi and CoCrFeMnNi) are listed in Table 2. $Q_{gb}$ and $Q_{gb}^*$ measured for Ni GB diffusion in the CoCrFeNi alloy are similar to the values for pure Ni and the Fe-Ni alloy. However, the frequency factor of Ni

GB diffusion, $P_0$, of CoCrFeNi is lower by an order of magnitude than that for Ni ($P_0^{CoCrFeNi}/P_0^{Ni} = 0.14$, Table 2). This fact may indicate the presence of some short-range order in the CoCrFeNi alloy, a fact which has also been pointed out earlier [10,35]. The quinary CoCrFeMnNi alloy, although showing a higher $P_0$ value than other systems, possesses the largest value of the activation energy and of the normalized activation energy for both, GB and bulk tracer diffusion of Ni. The enhanced energy barriers of bulk diffusion of Ni had been attributed to an increase in the average negative enthalpy of mixing of Ni with other constituents upon Mn addition [10]. A similar trend can be anticipated for GB diffusion as well.

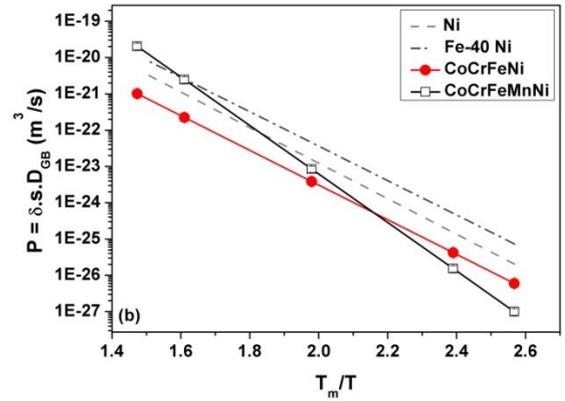

**Fig 8**. Arrhenius plots for Ni GB diffusion in CoCrFeNi and CoCrFeMnNi alloys and comparison of GB diffusivities (triple products) of Ni in pure Ni [27] and in Fe-Ni alloy [32] against the inverse of the homologous temperature

Another similarity between bulk and GB diffusion of Ni in these HEAs is the appearance of a cross-over temperature ($T_c$ = 800 K for GB diffusion and 970 K for bulk diffusion, see Ref. [10]). Below $T_c$, diffusion in the CoCrFeMnNi alloy is slower than in the CoCrFeNi alloy, while the reverse trend is observed above $T_c$. As explained in [10], $T_c$ arises due to two opposite effects of the Mn addition − (i) an increase of the average negative enthalpy of mixing $(\Delta H^{Ni}_{avg}$ (CoCrFeNi) = −3 kJ/mol; $\Delta H^{Ni}_{avg}$ (CoCrFeMnNi) = −4.25 kJ/mol) and (ii) a decrease of the melting temperature of the alloy upon Mn alloying [10]. A careful look at Fig. 6b suggests that a cross-over temperature ($T'_c$ = 1072 K) also exists between the Fe−40Ni alloy and CoCrFeMnNi. The much higher value of $T'_c$ as compared to $T_c$ is in line with a larger difference in the average enthalpy of mixing for Ni between the binary and quinary alloys ($\Delta H^{Ni}_{avg}$ (Fe–Ni) = −2 kJ/mol; $\Delta H^{Ni}_{avg}$ (CoCrFeMnNi) = −4.25 kJ/mol). The resemblance of trends between bulk and GB diffusion should not be treated as obvious, as GB diffusion in materials may not be always be controlled by thermodynamic factors and can be sensitive to various other aspects such as the level of purity [24], GB motion [36], solute segregation [37], and the particular GB structure [38] or the presence of GB-complexions [39].

*Grain boundary energy*

An important parameter which is critical for such phenomena as wetting, corrosion, microstructure stability, etc., is the grain boundary energy, $\gamma_{gb}$. It is also a valuable input for first-principle calculations predicting, e.g., nucleation-related processes at GBs. Borisov *et al.* [40] proposed a semi-empirical relation between the grain boundary energy and the self-diffusion parameters in pure metals and alloys, which after Gupta [41] is given as:

$$\gamma_{gb} = \frac{RT}{2a_0^2 N_A} \ln\left(\frac{D_{0\,gb}}{D_{0\,v}}\right) + \frac{1}{2a_0^2 N_A}(Q_v - Q_{gb}) \quad (6)$$

Here $a_0$ is the lattice parameter and $N_A$ the Avogadro's number. Guiraldenq [42] extended this concept to binary alloys and based on the reasoning in Ref. [24] we propose to apply this relation to HEAs, too. Using $a_0 = 0.355$ nm for both alloys and taking the bulk tracer diffusion parameters of Ni diffusion in HEAs from our previous study [10] and the GB diffusion values obtained in this work, the grain boundary energies of CoCrFeNi and CoCrFeMnNi are evaluated as functions of temperature,

$\gamma_{gb} = 0.632 + (1.55 \times 10^{-4}) \cdot T$ for CoCrFeNi (7)

and

$\gamma_{gb} = 0.553 + (3.37 \times 10^{-4}) \cdot T$ for CoCrFeMnNi (8)

These relations are presented in Fig. 9 and compared with the grain boundary energies determined for Ni of different purity levels using the same approach. The positive value of the temperature coefficients in above expressions indicates that the residual impurities present in commercially pure alloys segregate to grain boundaries at lower temperatures causing a decrease of the grain boundary energy [27]. Note that in [43] a decrease of the GB energy with temperature was reported for high-purity copper, 99.9998wt.%, whereas its increase in a less pure material, 99.999wt.%, was attributed to strong sulphur segregation. The same trend is seen in Fig. 9, the GB energy decreases with temperature in high-purity Ni and an opposite tendency is seen for Ni of lower purity levels. In a pure material, a decrease of the GB energy with temperature is generally expected from simple thermodynamic reasons, and the opposite tendency in an alloy is attributed to elemental segregation to the grain boundaries [44]. However, our atom-probe tomography measurements do not reveal any substantial element segregation (including that of residual carbon) to the arbitrarily chosen grain boundaries in the HEAs under investigation. A systematic investigation, especially on fully characterized bicrystals, is required to shed further light into this intriguing behavior.

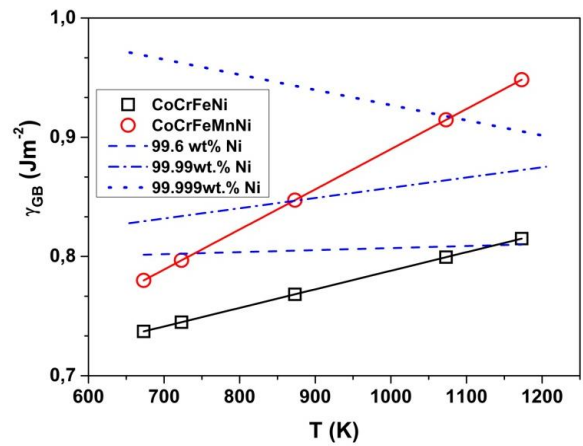

**Fig. 9:** Variation of GB energy, $\gamma_{gb}$, with temperature T for pure Ni of different purity levels (99.6 wt.% [27], 99.99 wt.% [24], 99.999 wt.% [27]) and HEAs (symbols, the present work).

The higher value of the GB energy for the quinary alloy, compared to the quaternary alloy can possibly arise due to a higher negative enthalpy of mixing of the former. This fact also indicates a higher driving force for nucleation at GBs in CoCrFeMnNi that may probably facilitate the phase decomposition in the quinary alloy with respect to that in the quaternary one. The temperature coefficients decrease from CoCrFeMnNi to CoCrFeNi and furthermore to pure Ni. A detailed investigation of the structure and properties of GBs in these HEAs would be required to explain these observations, which is beyond the scope of the present study. However, the GB energies calculated from the results of the diffusion measurements are in a good agreement with the experimentally determined values in coarse grained Ni [24]. Thus, in absence of any experimental reports on GB energies in HEAs to the best of our knowledge, our results offer an attractive set of reliable data for the studied temperature range. The determined strong temperature dependence of the GB energies in HEAs needs a further study, since the exact value of the GB energy is important for the phase stability and the decomposition of the solid solution at grain boundaries.

**Conclusions**

Detailed knowledge of diffusion along grain boundaries is fundamental to understand phase stability and creep deformation of high entropy alloys. For the first time, GB self-diffusion of Ni is measured in both CoCrFeNi and CoCrFeMnNi HEAs using the radiotracer technique. The chemical homogeneity of the alloys and the absence of elemental segregation at GBs have been confirmed using APT.

CoCrFeMnNi shows a lower GB diffusivity than CoCrFeNi only below 800 K, the tendency, however, is reversed at higher temperatures. This is due to two opposite effects upon Mn addition, namely more negative $\Delta H^{Ni}_{avg}$ and a decrease of the melting temperature. It conclusively shows that the presumption of decreasing diffusivities in HEAs merely due to an increase of the number of elements does not hold. The nature of added components plays a more decisive role. However, the chemically disordered nature of the HEA matrix does have its influence in reducing the diffusion rates in HEAs when compared to pure Ni and binary alloys, particularly at lower temperatures ($D_{Ni}/D_{CoCrFeMnNi} \approx 2$ at 800 K). Apparently, multi-element surroundings cause an increased energy barrier in CoCrFeMnNi and reduced diffusion entropy contributions (hence lower pre-exponential factor) in CoCrFeNi for the diffusing atom (Ni). This further suggests that decreased diffusion in HEAs is not a function of configurational entropy alone and several atomistic factors can contribute depending on the HEA considered.

The reliability of radiotracer measurements also allow us to derive GB energy values using the GB diffusion coefficients obtained. The grain boundary energies are determined as a function of temperature in both alloys and large positive temperature coefficients are found indicating that the GB energy decreases significantly with decreasing temperature.

**Acknowledgements**

Financial support from the German Science Foundation, DFG, project DI 1419/13-1, is acknowledged. M. Vaidya would also like to thank DAAD for providing a scholarship to carry out tracer diffusion experiments at University of Muenster, Germany. The authors also acknowledge the support of DAAD in facilitating the RWTH-IITM strategic partnership program.